\documentclass[aps,pre,onecolumn,showpacs]{revtex4}
\usepackage{graphicx}
\usepackage{epsfig}
\usepackage{wasysym}
\newcommand{\be}{\begin{equation}}
\newcommand{\ee}{\end{equation}}
  
\begin{document}
  
\title{Eddy current damping of a moving domain wall: beyond the
quasistatic approximation} 

\author{Francesca Colaiori$^1$, Gianfranco
Durin$^2$, and Stefano Zapperi$^1$} 
\affiliation{$^1$CNR-INFM, SMC,
Dipartimento di Fisica, Sapienza Universit\`a di Roma, P.le A. Moro 2,
00185 Roma, Italy} 
\affiliation{$^2$ INRIM, str. delle Cacce 91, 10137
Torino, Italy}
\begin{abstract}
In conducting ferromagnetic materials, a moving domain wall induces
eddy currents in the sample which give rise to an effective retarding
pressure on the domain wall. We show here that the pressure is not
just proportional to the instantaneous velocity of the wall, as often
assumed in domain wall models, but depends on the history of the
motion.  We calculate the retarding pressure by solving the Maxwell
equations for the field generated by the eddy currents, and show how
its effect can be accounted for by associating a negative effective
mass to the magnetic wall. We analyze the dependence of this effect on
the sample geometry and discuss the implications for Barkhausen noise
measurements.
\end{abstract}
\pacs{75:60.Ch, 75:60.Ej}
\maketitle

\section{Introduction}

Soft magnetic materials subject to a slowly changing external magnetic
field, respond with a jerky motion of the domain walls, known as
Barkhausen noise \cite{review, bertotti}. Many universal properties of
this noise have been identified, and are correctly reproduced by
theoretical models, confirming that, in analogy to critical phenomena,
the statistical features of the signal only depend on general
properties of the physical mechanism that govern the magnetization
reversal process, while they are independent from the microscopic
details of the particular system \cite{ABBM1,ABBM2,weiss}.

The average shape of the pulse, being a non--scalar quantity, has been
successfully identified as a powerful tool to characterize the
universal properties of crackling systems: pulses of different
duration, once properly rescaled, are expected to collapse onto a
universal function \cite{sethna}.  In the case of Barkhausen noise,
pulses from experimental data do approximately collapse on the same
curve \cite{asym,metha}, however, this curve shows a clear leftward
asymmetry, while models that very accurately reproduces most of the
other universal quantities predict a symmetric shape \cite{shape}.

As shown in Ref.~\cite{nature}, the asymmetry in Barkhausen pulses is
due to the non--instantaneous response of the eddy current field to
the domain wall displacement.  Barkhausen noise models usually assume
this response to be instantaneous, and thus do not capture this
asymmetry. Since eddy currents take a finite time to set up and also
they persist for a finite time after the corresponding wall
displacement, the pressure on the moving domain wall is not strictly
proportional to the instantaneous velocity of the wall, but depends on
the history of the motion. The delay has a characteristic timescale,
and therefore its effect is more evident on avalanches of comparable
duration, and disappears on very large ones, where the separation of
timescales is such that the response of the field can be assumed to be
instantaneous, and strict universality is recovered. The first order
correction to the instantaneous response approximation can be
accounted for by associating a negative effective mass to the wall in
the equation of motion \cite{nature}.
 
In this paper we report a detailed calculation of the retarding
pressure starting from Maxwell equations, and obtain the negative
effective mass as a first order correction to the quasi--static
approximation. Our approach is similar to the one of
Bishop\cite{bishop}. The resulting non-local damping was previously
employed in a domain wall dynamics model in Ref.~\cite{nature}, where
its effect on the Barkhausen pulse shape was studied and compared with
experimental data. Here, we analyze the role of sample geometry on the
eddy current propagation and provide expression for the damping term
and the effective mass as a function of the sample aspect ratio.

The paper is organized as follows. In Sec. II we solve the Maxwell
equations for the eddy currents in a conducting sample with a moving
domain wall. In Sec. III we compute the resulting pressure on the
domain wall.  In Sec. IV we derive the first order correction to the
pressure. In Sec. V we discuss the role of the sample geometry and in
Sec. VI we conclude. Finally, two appendix report the details of some
series summation used in the manuscript.

\section{The eddy current field from the Maxwell equations}
\begin{figure}[h]
\centerline{\psfig{file=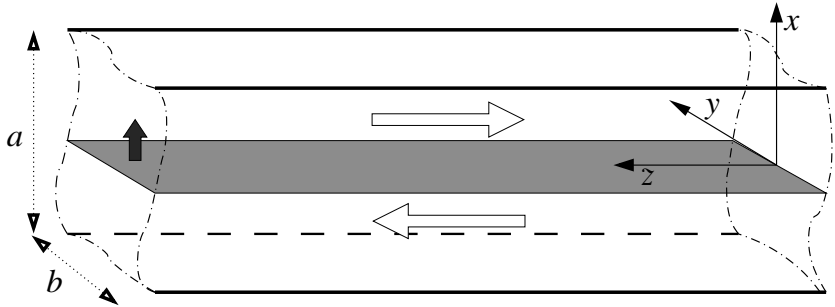,width=11cm,clip=!}}
\caption{Horizontal arrows indicate the directions of the
  magnetization in the two domains. The black vertical arrow indicates the
  direction of motion of the wall.}
\label{geometry}
\end{figure}

Consider a sample with dimension $x\in[-a/2,a/2]$,
$y\in[-b/2,b/2]$, and infinite in the $z$ direction, divided in two
magnetic domains by a rigid domain wall on the $yz$ plane, moving from
position $x=0$, as in Fig.\ref{geometry}.  The displacement of a
magnetic wall in a conducting medium induces a flow of eddy currents
that generates a magnetic field, which, in the geometry indicated in
Fig. \ref{geometry}, is parallel to the $z$ axis:
\begin{equation}
\vec{H}= H(x,y,t)\hat{z}\,.
\label{H}
\end{equation}
The magnetic field obeys the Maxwell equation
\begin{equation}
\nabla^2H=\sigma \mu \partial_tH\,,
\label{Maxwell}
\end{equation}
where $\sigma$ and $\mu$ are the electric conductivity and the
magnetic permeability of the medium. In Eq. \ref{Maxwell} the
displacement currents are neglected with respect to the ohmic
currents.  This equation is usually solved in the quasi--static
approximation, where $\mu$ is negligible within domains, and the
equation reduces to $\nabla^2H=0$.  Eq. \ref{Maxwell} is a diffusion
equation: there is a finite time delay between the wall displacement,
and the establishment of the eddy currents. Given that the typical
timescale for diffusion is proportional to $\mu$, the quasi--static
approximation $\mu=0$ corresponds to assuming an instantaneous
response of the field.

To take into account dynamical effects from eddy currents in the
domain wall propagation, we need to solve Eq.  \ref{Maxwell}, with the
appropriate boundary condition $H=0$ on the sample surface. The
discontinuity across the wall is regulated by the Faraday condition
$\partial_xH(0^+,y,t)-\partial_xH(0^-,y,t) =2\sigma I v(t)$, where
$v(t)$ is the velocity of the wall.

Let us expand $H$ in its Fourier components
\begin{equation}
H(x,y,t)=\frac{1}{\sqrt{2\pi}}\int_{-\infty}^{\infty} d\omega
F(x,y,\omega)
e^{i\omega t}\,.
\label{H2}
\end{equation} 
Each component has to satisfy 
\begin{equation}
\nabla F(x,y,\omega)=r^2F(x,y,\omega)\,, 
\label{F}
\end{equation}
with $r^2=i\omega \mu \sigma$. The magnetic field has to be zero on
the sample boundary, which implies the conditions $F(x,\pm
b/2,\omega)=F(\pm a/2,y,\omega)=0$.  Moreover, the Faraday condition
around the wall imposes
$\partial_xF(0^+,y,\omega)-\partial_xF(0^-,y,\omega) =2\sigma
I\hat{v}(\omega),$ where $I$ is the saturation magnetization, and
$\hat{v}$ is the Fourier transform of the velocity of the wall $v(t)$.
The solution is of the form
\begin{equation}
F(x,y,\omega)=\sum_{n=0}^{\infty}A_n(x,\omega) \cos(\lambda_n y)\,,
\label{F2}
\end{equation}
where $A_n$ satisfies
\begin{equation}
\partial^2_{x}A_n(x,\omega)=\Lambda_n^2 A_n(x,\omega)\,,
\label{A}
\end{equation}
with $\Lambda_n^2=\lambda^2+r^2$, to be solved separately for $x>0$
and $x<0$. The condition on the sample boundary in the $y$ direction
implies $\cos(\pm b/2\lambda_n)=0$ which fixes
$\lambda_n=(2n+1)\pi/b$.  The condition on the other boundary in
satisfied by choosing
\begin{equation}
A_n(x,\omega)=C_n(\omega) \sinh(\Lambda_n(\mid x \mid - a/2))\,,
\label{A2}
\end{equation}
so that 
\begin{equation}
F(x,y,\omega)=\sum_{n=0}^{\infty}C_n(\omega) 
\sinh(\Lambda_n(\omega)(\mid x \mid -a/2))\cos(\lambda_n y)\,.
\label{F3}
\end{equation}
The functions $C_n(\omega)$ are fixed by the Faraday condition
\begin{equation}
\partial_xF(0^+,y,\omega)=-\partial_x F(0^-,y,\omega)=
\sum_{n=0}^{\infty}C_n(\omega)
\Lambda_n(\omega) \cosh(\Lambda_n(\omega) a/2) 
\cos(\lambda_n y)=\sigma I \hat{v}(\omega)\,.
\label{}
\end{equation}
Multiplying by $\cos(\lambda_m y)$, integrating in $[-b/2,b/2]$, and
using the orthogonality relations $\int_{-b/2}^{b/2}dy \cos(\lambda_n
y) \cos(\lambda_m y)=\delta_{n,m}b/2$, and $\int_{-b/2}^{b/2}dy
\cos(\lambda_m y)=(-1)^m2/\lambda_m$ we get
\begin{equation}
C_n(\omega)=
(-1)^n\frac{4\sigma I}{b} \frac{1}{\lambda_n
  \Lambda_n(\omega) \cosh(\Lambda_n(\omega) a/2)}\hat{v}(\omega)\,,
\label{C}
\end{equation}
so that finally
\begin{equation}
F(x,y,\omega)=
\sum_{n=0}^{\infty}(-1)^n
\frac{4\sigma  I}{b}\frac{\tanh(\Lambda_n(\omega) a/2)}
{\lambda_n  \Lambda_n(\omega) }\cos(\lambda_n y)\hat{v}(\omega)\,.
\label{F4}
\end{equation}

\section{Eddy current pressure on the wall}
Once the solution of Maxwell equation with the appropriate boundary
condition is given, the average eddy current pressure on the wall is
obtained by integrating the magnetic field over $y$ at the wall
position $x=0$:
\begin{equation}
P(t)=\frac{2I}{b}\int_{-b/2}^{b/2}dy
H_z(0,y,t)=\frac{2I}{b}
\frac{1}{\sqrt{2\pi}}
\int_{-\infty}^{\infty} d\omega e^{i \omega t} 
\int_{-b/2}^{b/2}dy F_{\omega}(0,y)\,,
\label{Pt}
\end{equation}
or, in terms of the Fourier transform: 
\begin{equation}
\hat{P}(\omega)=-\hat{v}(\omega)\hat{f}(\omega)\,,
\label{Po2}
\end{equation}
with
\begin{equation}
\hat{f}(\omega)=\frac{16 I^2
  \sigma}{b^2}\sum_{n=0}^{\infty}\frac{\tanh(\Lambda_n(\omega)
  a/2)}{\lambda_n^2 \Lambda_n(\omega)} \,.
\label{fo}
\end{equation}
In real space the pressure at time $t$ is given by a convolution of
velocities of the wall at all times prior to $t$ with the response
function $f$: 
\begin{equation}
P(t)=\frac{1}{\sqrt{2\pi}}\int_{-\infty}^{\infty} ds v(t-s) f(s)
\label{Pt22}\,.
\end{equation}
To simplify the $\omega$ dependence in Eq. \ref{fo} we use the
relation  
\begin{equation}
\tanh(z)=8z\sum_{k=0}^{\infty}\frac{1}{\pi^2 (2k+1)^2 +4 z^2}\,,
\label{trick}
\end{equation}
with $z=\Lambda_n(\omega) a/2$, which gives
\begin{equation}
\hat{f}(\omega)=
\frac{64 a I^2 \sigma}{b^2}
\sum_{n,k=0}^{\infty}\frac{1}{\lambda_n^2 
(b^2 \lambda_k^2+a^2 \Lambda_n^2(\omega))}\,.
\label{ft}
\end{equation}
Replacing $\lambda_n$ and $\Lambda_n$ with their expressions, we get 
\begin{equation}
\hat{f}(\omega)=\frac{64 I^2}{a b^2 \sigma \mu^2}
\overline{\sum_{n,k=1}^{\infty}}\frac{1}
{n^2\omega_b\left(k^2\omega_a+n^2\omega_b+i\omega\right)}\,,
\label{fo2}
\end{equation}
where $\overline{\sum}$ indicates a summation over odd numbers only,
and $\omega_a=\tau_a^{-1}=\pi^2/\sigma \mu a^2$,
$\omega_b=\tau_b^{-1}=\pi^2/\sigma \mu b^2$.  Given that
$1/(\omega_0+i\omega)$ has $\sqrt{2\pi} \exp(-\omega_0 t) \theta(t)$
as inverse Fourier transform (where $\theta(t)$ is the Heaviside theta
function), it is easy to obtain from Eq. \ref{fo2} the expression for
$f$ in real space by anti--transforming term to term in the double sum
to get
\begin{equation}
f(t)=\sqrt{2\pi}\frac{64 I^2}{a b^2 \sigma \mu^2}
\overline{\sum_{n,k=1}^{\infty}}\frac{1}{n^2\omega_b}
e^{-\omega_{k,n}t}\theta(t)\,.
\label{f}
\end{equation}
where $\omega_{k,n}=\tau_{k,n}^{-1}=k^2\omega_a+n^2\omega_b$.  The
response function results to be the sum of simple exponential
relaxations, with different relaxation times. The largest and
therefore most relevant relaxation time is
$\tau_{0,0}=\frac{\sigma\mu}{\pi^2}\left(\frac{1}{a^2}+\frac{1}{b^2}\right)^{-1}$.

The form \ref{f} of the response function gives through the
convolution \ref{Pt} the explicit solution for the retarded pressure
on the wall.

\section{First order correction: Damping coefficient and negative effective mass}
Given the full solution of the response function, we now want to
calculate the first order correction to the quasi--static
approximation $\mu=0$ in the retarded pressure.  In order to do this,
let us replace the expression \ref{f} in the convolution \ref{Pt} and
exchange the sum with the integral:
\begin{equation}
P(t)=\frac{64 I^2}{a b^2 \sigma \mu^2}
\overline{\sum_{n,k=1}^{\infty}}\frac{1}{n^2\omega_b}
\int_0^{\infty} ds v(t-s) e^{-\omega_{k,n}s} \,,
\label{Pt2}
\end{equation}
where the Heaviside function was eliminated by restricting the domain
of integration. For small relaxation times (note that
$\tau_{k,n}<\tau_{0,0}$ for every $k$ and $n$) the exponential
functions in the integral decay very fast around $s=0$, so that the
velocities that sensibly contribute to the convolutions are only those
at time very close to $t$. We may thus expand $v(t-s)\simeq
v(t)-sv'(t)$ around $s=t$, and perform the integrals. The term
proportional to $v(t)$ corresponds the usual instantaneous
contribution, while the term proportional to $v'(t)$ gives rise to the
first order correction to the quasi--static approximation:
\begin{equation}
P(t)\simeq\frac{64 I^2}{a b^2 \sigma \mu^2}
\overline{\sum_{n,k=1}^{\infty}}\frac{1}{n^2\omega_b}
\left(
\frac{v(t)}{\omega_{k,n}}-\frac{v'(t)}{\omega_{k,n}^2}
\right)\,.
\label{Pt3}
\end{equation}
Replacing the frequencies with their expression
\begin{equation}
P(t)\simeq\frac{64 I^2\sigma}{\pi^4 }\frac{b^2}{a}
\left(v(t)
\Sigma_1(b/a)-v'(t)\tau_b\Sigma_2(b/a)
\right)\,,
\label{Pt4}
\end{equation}
where $\Sigma_1(\alpha)= \overline{\sum}_{n,k} \,\frac{1}{n^2}
\frac{1}{n^2+\alpha^2 k^2}$, and $\Sigma_2(\alpha)=
\overline{\sum}_{n,k} \,\frac{1}{n^2} \frac{1}{(n^2+\alpha^2 k^2)^2}$.
Eq. \ref{Pt4} allows to identify a damping coefficient
\begin{equation}
\Gamma=\frac{64 I^2\sigma b^2}{a \pi^4 } \Sigma_1(b/a)\,,
\label{damping}
\end{equation}
and an effective mass
\begin{equation}
M=-\frac{64 I^2 \sigma^2 \mu b^4}{a \pi^6} \Sigma_2(b/a)\,,
\label{mass}
\end{equation}
which turns out to be negative.  Moreover, a characteristic time
$\tau$ of order $\mu$ can be identified as the ratio between mass and
damping:
\begin{equation}
\tau= |M|/\Gamma=\tau_b \frac{\Sigma_2(b/a)}{\Sigma_1(b/a)}\,.
\label{tau}
\end{equation}

A similar calculation to get the damping coefficient and the effective
mass may be carried out in frequency space. From Eq. \ref{fo2}, that
gives the Fourier transform of the full response function, we can
separate the real and the imaginary part to get
\begin{equation}
\hat{f}(\omega)=\frac{64 I^2}{a b^2 \sigma \mu^2}
\overline{\sum_{n,k=1}^{\infty}}\frac{k^2\omega_a+n^2\omega_b-i\omega}
{n^2\omega_b\left((k^2\omega_a+n^2\omega_b)^2+\omega^2\right)}
\,,
\label{fo3}
\end{equation}
which allows, by writing $\hat{f}(\omega)=\Gamma(\omega)-i \omega M$,
to formally identify a frequency dependent damping coefficient
\begin{equation}
\Gamma(\omega)=\frac{64 I^2\sigma b^2}{a \pi^4 } 
\overline{\sum_{n,k=1}^{\infty}}\frac{k^2 (a/b)^2+n^2}
{n^2\left((k^2(a/b)^2+n^2)^2+(\omega/\omega_b)^2\right)}\,,
\label{damping_omega}
\end{equation}
and effective mass 
\begin{equation}
M(\omega)=-\frac{64 I^2 \sigma^2 \mu b^4}{a \pi^6} 
\overline{\sum_{n,k=1}^{\infty}}\frac{1}
{n^2\left((k^2(a/b)^2+n^2)^2+(\omega/\omega_b)^2\right)}\,.
\label{mass_omega}
\end{equation}
Expanding to the first order in $\mu$
\begin{equation}
\hat{f}(\omega)\simeq
\frac{64 I^2 \sigma b^2}{a \pi^4}
\Sigma_1(b/a)
-i\omega \frac{64 I^2 \sigma^2 \mu b^4}{a \pi^6}
\Sigma_2(b/a)
\,,
\label{fo4}
\end{equation}
we recover $\Gamma$ and $M$ as in Eqs. \ref{damping},\ref{mass}.

\section{Dependence of damping and mass on the geometry of the sample}
Eqs. \ref{mass},\ref{damping},\ref{tau} give the expression correct to
the first order in $\mu$ of damping coefficient, effective mass and
characteristic time in a general geometry, defined by the parameters
$a$ and $b$.  The series $\Sigma_1$ and $\Sigma_2$ can be summed up
for particular geometries. The calculations are reported in the
appendixes.

For a slab with $a\gg b$  
\begin{equation}
\Sigma_1(b/a\rightarrow 0)\simeq \frac{a}{b}\frac{\pi}{4} \lambda_3 \,\,,\,\,
\Sigma_2(b/a\rightarrow 0)\simeq \frac{a}{b}\frac{\pi}{8} \lambda_5 \,,\,\,
\end{equation} 
where $\lambda_n=\overline{\sum}_k k^{-n}$, with $\lambda_3=1.05179...$
and $\lambda_5=1.00452...$,   
so that 
\begin{equation}
\Gamma=I^2 \sigma \frac{16\lambda_3}{\pi^3} b\,\,,\,\,
M=-I^2 \sigma^2 \mu \frac{8\lambda_5}{\pi^5}b^3\,\,,\,\,
\tau=\sigma \mu \frac{\lambda_5}{2 \pi^2 \lambda_3}b^2\,.
\label{ainft}
\end{equation}

For a slab with $a\ll b$  
\begin{equation}
\Sigma_1(b/a\rightarrow \infty)\simeq  \frac{\pi^4}{64}\frac{a^2}{b^2}\,\,,\,\,
\Sigma_2(b/a\rightarrow \infty)\simeq \frac{\pi^6}{768}\frac{a^4}{b^4}\,,\,\,
\end{equation}
so that 
\begin{equation}
\Gamma=I^2\sigma a\,\,,\,\,
M=-I^2\sigma^2\mu\frac{a^3}{12}\,\,,\,\,
\tau=\sigma\mu \frac{a^2}{12}\,.
\label{binft}
\end{equation} 

For a square rod with $a=b$ 
\begin{equation}
\Sigma_1(1)=\frac{\pi^4}{128} \,\,,\,\,
\Sigma_2(1)\simeq 0.264 \pm 0.001\,,
\end{equation}
so that
\begin{equation}
\Gamma=I^2 \sigma \frac{b}{2} \,\,,\,\,
M\simeq -I^2 \sigma^2 \mu \, 0.0175 \, b^3\,\,,\,\,
\tau \simeq \sigma \mu \, 0.0351 \, b^2\,.
\label{aeqb}
\end{equation}
\begin{table}[h]
\begin{tabular}{|l||c|c|c|}
\hline
               &$a\gg b$&$a\ll b$&$a=b$  \\
\hline
$\Gamma /(I^2 \sigma x)$&$0.054$&$ 1$ &$0.5$\\
$M /(I^2 \sigma^2\mu x^3)$&$0.026$&$0.083$&$0.0175$\\
$\tau /(\sigma \mu x^2)$&$0.048$&$0.083$&$0.035$\\
\hline 
\end{tabular}
\label{tab}
\caption{Values of $\Gamma$, M, and $\tau$ for varies geometries. $x$
  denotes the smaller between $a$ and $b$.}
\end{table}

The results are summarized in Tab. I.
The dependence on $a$ ($b$) correctly disappears in the limit
$a\rightarrow \infty$ ($b\rightarrow \infty$). The physical quantities
$\Gamma$, $M$, and $\tau$ all increase with the overall sample
size. For example, for a sample of a given thickness $a$, both damping
coefficient and effective mass increase with $b$ and $b^3$
respectively. However, as soon as $b$ becomes larger than $a$, they
saturate to a value proportional to $a$ and $a^3$ respectively. Since
the mass increases faster than the damping with $b$, the
characteristic time also increases with $b$, and saturates to a value
proportional to $a^2$. A similar behavior is observed by varying $a$
at fixed $b$, although the role of the two dimensions transverse and
parallel to the wall is not symmetric. Essentially the dependence of
$\Gamma$, $M$, and $\tau$ on the geometry of the sample is dominated
by the smaller between $a$ and $b$. Thus the relevance of the eddy
current dynamic effect, is controlled by the smaller sample dimension:
the thinner the sample, the smaller the effect, while the rest of the
geometry does not play any relevant role.  The characteristic time in
$\sigma \mu$ units is approximately equal to $\tau/\sigma \mu \simeq
0.083 \,a^2$, when $a<b$, $\tau/\sigma \mu \simeq 0.048 \,b^2$ when
$b<a$, and has its minimum $\tau/\sigma \mu \simeq 0.035 \,b^2$ for a
square rod. In all cases, it stays between $3$ and $10\%$ of the
squared relevant sample size.

\begin{figure}[h!b]
\centerline{\epsfig{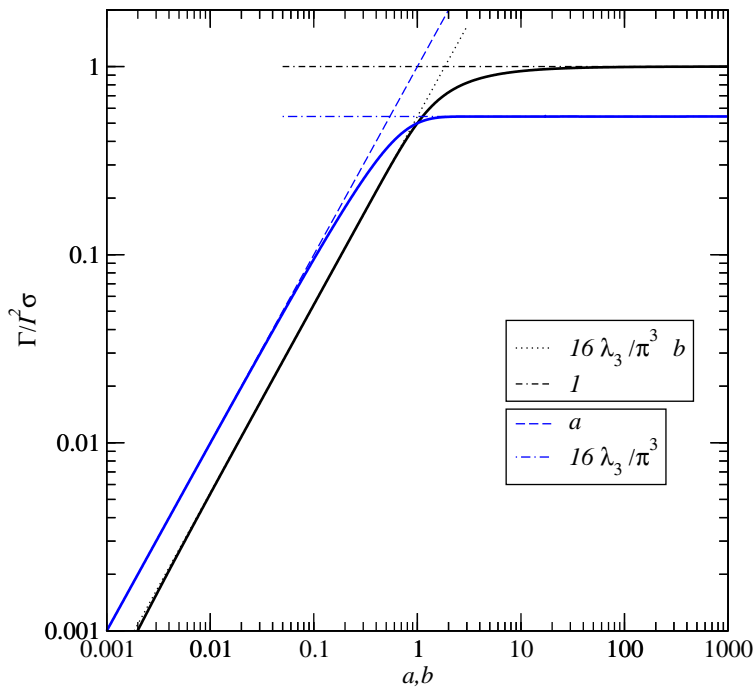}}
\caption{Log--log plot of the damping coefficient $\Gamma$ in $I^2
  \sigma$ units as a function of $b$ for $a=1$, and as a function of
  $a$ for $b=1$. The fits correspond to the analytic calculation.}
\label{damping.eps}
\end{figure}

\begin{figure}[h!t]
\centerline{\epsfig{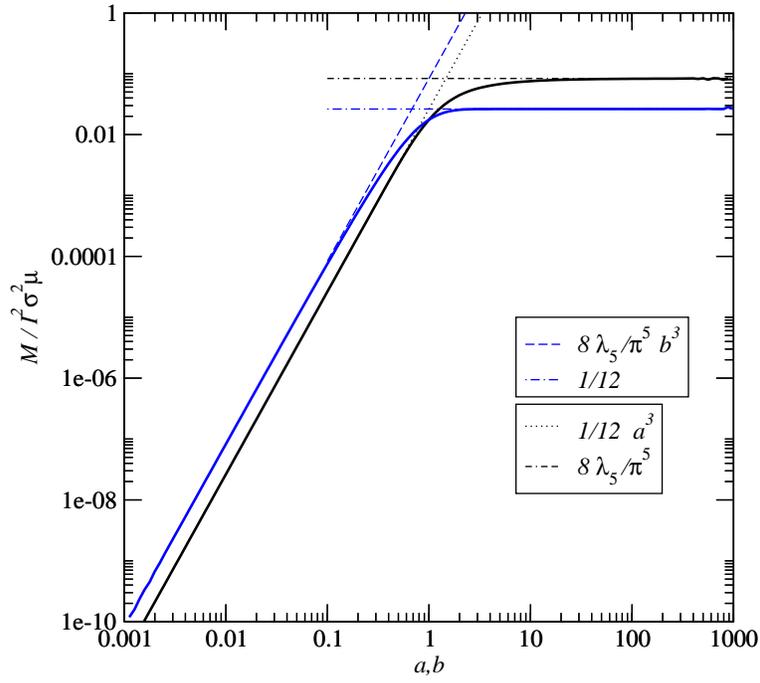}}
\caption{Log--log plot of the modulus of the effective mass $M$ in
  $I^2 \sigma^2 \mu$ units as a function of $b$ for $a=1$, and as a
  function of $a$ for $b=1$.  The fits correspond to the analytic
  calculation.}
\label{mass.eps}
\end{figure}

\begin{figure}[h!]
\centerline{\epsfig{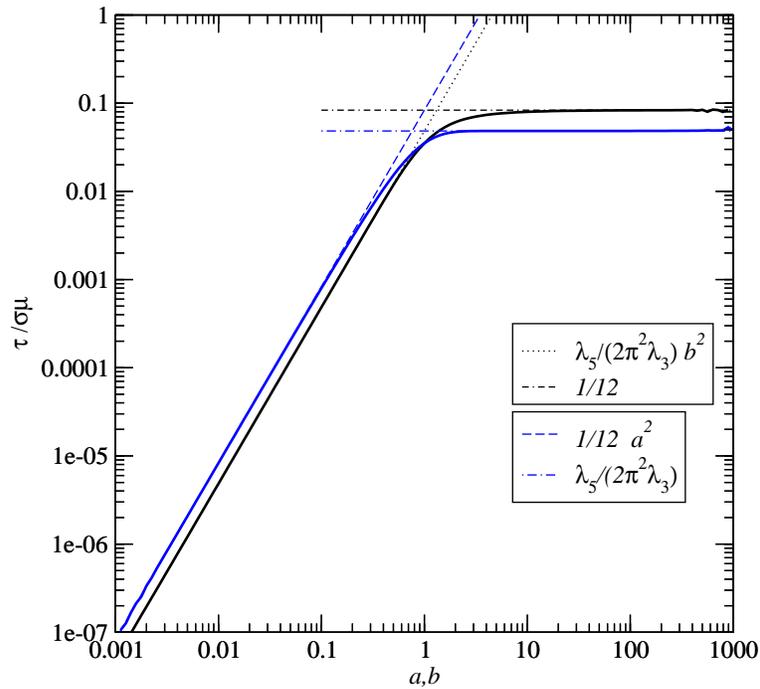}}
\caption{Log--log plot of the characteristic time $\tau$ in $\sigma
  \mu \sigma$ units as a function of $b$ for $a=1$, and as a function
  of $a$ for $b=1$. The fits correspond to the analytic calculation.}
\label{tau.eps}
\end{figure}

Figs. \ref{damping.eps},\ref{mass.eps}, and \ref{tau.eps} show the
behavior of $\Gamma$, $M$, and $\tau$ both as a function of parallel
dimension $b$, for $a=1$, and of the transvers dimension $a$, for
$b=1$. The straight lines are fits with the asymptotic behaviors
calculated for the large $a$ and $b$ limits.

\vspace{1.5cm}

\vspace{3.5cm}

\vspace{1.5cm}
$\mbox{}$
\vspace{1.5cm}
\vspace{4.5cm}

\section{Comments and conclusions}

The asymmetry in Barkhausen pulses is due to the non--instantaneous
response of the field to the domain wall displacement.  Most
treatments of Barkhausen noise assume the eddy current drag on the
magnetic domain wall to be instantaneous, which corresponds to
assuming $\mu=0$ within the magnetic domains. However, eddy currents
take a finite time to set up after the magnetic reversal, and persist
for a finite time after the corresponding wall displacement. This time
delay causes the eddy pressure on the wall at a given time $t$ not to
be strictly proportional to the instantaneous velocity of the wall,
but to depend on a weighted average of velocities of the wall up to
time $t$.  This quasi--inertial effect is responsible for the
asymmetry in the pulse shape observed in Barkhausen experiments.  The
delay has a characteristic timescale $\tau$, and therefore its effect
is more severe on avalanches of comparable duration. For very long
avalanches the separation of timescales is such that the response of
the field can be assumed to be instantaneous, therefore the asymmetry
disappears, and strict universality is recovered.

Starting from the full Maxwell equation, that includes the dynamic
eddy current effects, we calculate the retarded pressure on the wall.
The first order correction to the quasi--static ($\mu=0$) solution,
leads to identify a damping coefficient, which survives to the $\mu
\rightarrow 0$ limit and coincides with the one calculated assuming an
instantaneous response, and an effective mass, that vanishes in the
$\mu\rightarrow 0$ limit.  A damping coefficient and an effective mass
can be formally defined beyond the first order approximation, however,
they both results to be frequency dependent.  Damping coefficient,
effective mass, and characteristic time depend on the sample
geometry, however, it turns out that the only geometrical parameter
that significantly affects them is the smaller sample dimension.

The effective mass results to be negative at all frequencies.  This
may be understood by observing that the retarded pressure at time $t$
is proportional to an average of previous velocities of the wall up to
time $t$. When the wall is accelerating, the effective average
velocity is smaller than the instantaneous one, and the opposite is
true when the wall decelerates.  The leftward pulse asymmetry observed
in Barkhausen experiments is indeed consistent with a negative
effective mass: the avalanche start fast, and end slowly, which is
exactly the opposite of what one would expect from standard inertia.

\appendix
\section{Sum of the series $\Sigma_1(\alpha)$ for
  $\alpha=1$ and in the limit $\alpha\rightarrow\infty$,
  $\alpha\rightarrow 0$} 
We here calculate the sum of the series
\begin{equation}
\Sigma_1(\alpha)=\overline{\sum_{n,k=1}^{\infty}}\frac{1}{n^2}
\frac{1}{n^2+\alpha^2 k^2}
\end{equation}
for some specific values of $\alpha$. The value of $\Sigma_1$ for
$\alpha=1$ can be obtained by writing a closed equation for $\Sigma_1(1)$
as follows:
\begin{equation}
\overline{\sum_{n,k=1}^{\infty}}
\frac{1}{n^2}\frac{1}{n^2+k^2}=
\overline{\sum_{n,k=1}^{\infty}}
\left(\frac{1}{n^2}-\frac{1}{n^2+k^2}\right)\frac{1}{k^2}=
\overline{\sum_{n,k=1}^{\infty}}
\frac{1}{n^2}\frac{1}{k^2}-
\overline{\sum_{n,k=1}^{\infty}}
\frac{1}{k^2}\frac{1}{n^2+k^2}
\end{equation}
which gives
\begin{equation}
2\overline{\sum_{n,k=1}^{\infty}}\frac{1}{n^2}
\frac{1}{n^2+k^2}=
\left(\overline{\sum_{n,k=1}^{\infty}}
\frac{1}{k^2}\right)^2=\left(\frac{\pi^2}{8}\right)^2
\end{equation}
so that 
\begin{equation}
\Sigma_1(1)=\pi^4/128 \,.
\end{equation}
To calculate $\Sigma_1(\alpha)$ in the limit $\alpha \rightarrow
0$ we use Eq. \ref{trick} with $z=n\pi/2\alpha$ to write
\begin{equation}
\overline{\sum_{n,k=1}^{\infty}}\frac{1}{n^2}
\frac{1}{n^2+\alpha^2 k^2}=
\frac{\pi}{4\alpha}
\overline{\sum_{n=1}^{\infty}}\frac{\tanh(n\pi/2\alpha)}{n^3}
\label{tanh}
\end{equation}
which, in the limit $\alpha \rightarrow 0$ gives 
\begin{equation}
\Sigma_1(\alpha\rightarrow 0)\simeq \frac{\pi}{4\alpha}\lambda_3 \,.
\end{equation}

In the opposite limit $\alpha \rightarrow \infty$, using again
Eq. \ref{tanh} and the sum $\lambda_n=\overline{\sum}_k
k^{-2}=\pi^2/8$ we get 
\begin{equation}
\Sigma_1(\alpha\rightarrow 0)\simeq \frac{\pi^4}{64\alpha^2} \,.
\end{equation}

\section{Sum of the series $\Sigma_2(\alpha)$ in the limit $\alpha\rightarrow\infty$,
  $\alpha\rightarrow 0$}
We here calculate the sum of the series
\begin{equation}
\Sigma_2(\alpha)=\overline{\sum_{n,k=1}^{\infty}}\frac{1}{n^2}
\left(\frac{1}{n^2+\alpha^2 k^2}\right)^2
\end{equation}
for some specific values of $\alpha$.
To calculate $\Sigma_1(\alpha)$ in the limit $\alpha \rightarrow
0$ let us write $\Sigma_2$ as 
\begin{equation}
\Sigma_2(\alpha)=\lim_{\epsilon\rightarrow 0}
\frac{1}{\epsilon}
\overline{\sum_{n,k=1}^{\infty}}\frac{1}{n^2}
\left(\frac{1}{n^2+\alpha^2 k^2}-\frac{1}{n^2+\alpha^2 k^2+\epsilon}\right)
\end{equation}
and then use Eq. \ref{trick} in both terms with $z=n\pi/2\alpha$ and
$z=\sqrt{n^2+\epsilon}\pi/2\alpha$ respectively
\begin{equation}
\Sigma_2(\alpha)=\lim_{\epsilon\rightarrow 0}
\frac{1}{\epsilon}
\overline{\sum_{n}^{\infty}}\frac{1}{n^2} 
\frac{\pi}{4\alpha}
\left(
\frac{\tanh\left(\pi n/2\alpha\right)}{n}-
\frac{\tanh\left(\pi \sqrt{n^2+\epsilon}/2\alpha\right)}
{\sqrt{n^2+\epsilon}}
\right)\,.
\end{equation}
Expanding for small $\epsilon$ and taking the limit one gets
\begin{equation}
\Sigma_2(\alpha)=
\frac{\pi}{8\alpha}\overline{\sum_{n}^{\infty}}
\frac{\tanh\left(\pi n/2\alpha\right)}{n^5}-
\frac{\pi^2}{16\alpha^2}\overline{\sum_{n}^{\infty}}
\frac{1-\tanh^2\left(\pi n/2\alpha\right)}{n^4}\,.
\label{S22}
\end{equation}
Now we can finally take the limit $\alpha \rightarrow 0$ to get
\begin{equation}
\Sigma_2(\alpha \rightarrow 0)\simeq 
\frac{\pi}{8\alpha}
\overline{\sum_{n}^{\infty}}\frac{1}{n^5}=
\frac{\pi \lambda_5}{8\alpha}  \,.
\end{equation}
The limit $\alpha \rightarrow \infty$ can also be obtained from 
Eq. \ref{S22} by using the expansion $\tanh(x)\simeq x-x^3/3$ for
small arguments, which gives
\begin{equation}
\Sigma_2(\alpha \rightarrow \infty)\simeq 
\frac{\pi}{8\alpha}\overline{\sum_{n}^{\infty}}\left(\frac{\pi}{2\alpha}\frac{1}{n^4}-
\frac{\pi^3}{3(8\alpha)^3}\frac{1}{n^2}
\right)-
\frac{\pi^2}{16\alpha^2}\overline{\sum_{n}^{\infty}}\left(
\frac{1}{n^4}-
\frac{\pi^2}{(2\alpha)^2}\frac{1}{n^2}
\right)
=
\frac{\pi^6}{768\alpha^4}  \,.
\end{equation}

\end{document}